\documentclass[conference]{IEEEtran}
\IEEEoverridecommandlockouts
\usepackage{cite}
\usepackage{amsmath,amssymb,amsfonts}
\usepackage{algorithmic}
\usepackage{graphicx}
\usepackage{textcomp}
\usepackage{subfig}
\def\BibTeX{{\rm B\kern-.05em{\sc i\kern-.025em b}\kern-.08em
    T\kern-.1667em\lower.7ex\hbox{E}\kern-.125emX}}
\begin{document}
\title{Decision Level Fusion: An Event Driven Approach
\thanks{Approved for Public Release: 18-MDA-9560 (22 March 18)}}

\author{\IEEEauthorblockN{Siddharth Roheda}
\IEEEauthorblockA{\textit{Department of Electrical and Computer Engineering} \\
\textit{North Carolina State University}\\
Raleigh, NC, USA \\
sroheda@ncsu.edu}
\and
\IEEEauthorblockN{Hamid Krim}
\IEEEauthorblockA{\textit{Department of Electrical and Computer Engineering} \\
\textit{North Carolina State University}\\
Raleigh, NC, USA\\
ahk@ncsu.edu}
\and
\IEEEauthorblockN{Zhi-Quan Luo}
\IEEEauthorblockA{\textit{Department of Electrical and Computer Engineering} \\
\textit{University of Minnesota}\\
Minneapolis, MN, USA\\
luozq@ece.umn.edu}
\and
\IEEEauthorblockN{Tianfu Wu}
\IEEEauthorblockA{\textit{Department of Electrical and Computer Engineering} \\
\textit{North Carolina State University}\\
Raleigh, NC, USA\\
tianfu\_wu@ncsu.edu}
}

\maketitle

\begin{abstract}
This paper presents a technique that combines the occurrence of certain events, as observed by different sensors, in order to detect and classify objects. This technique explores the extent of dependence between features being observed by the sensors, and generates more informed probability distributions over the events. 
Provided some additional information about the features of the object, this fusion technique can outperform other existing decision level fusion approaches that may not take into account the relationship between different features.
\end{abstract}

\begin{IEEEkeywords}
Sensor Fusion, Decision Level Fusion, Event based Classification, Coupling
\end{IEEEkeywords}

\section{Introduction}
\label{sec:intro}

One often resorts to fusing data from more than one sensors to successfully carry out target detection and classification. Data fusion has, over the years been recognized to distinguish three levels of fusion, namely data level, feature level, and decision level. Data level fusion generally processes raw data generated by each sensor, and performs the fusion of information according to some criterion before proceeding to inference. Feature level fusion, on the other hand, first gleans features from raw data (e.g., transformed data) from diverse sensors, to subsequently coherently merge them for inference. In decision level sensor fusion, each sensor reaches an individual  decision, prior to optimal combination of the decisions to yield a more informed decision, e.g. target classification.\par   
In this paper, we present on the decision level fusion strategy, and propose a principled approach to fusion for improved inference performance.  A classification decision is reached by cataloguing  sets of events along with their probabilistic characterization for each sensor, and following  a joint probabilistic and coherent evaluation of these events.  These events are formalized for each sensor according to its potentially extracted features to define different targets (e.g. target radar cross section less than $10m^2$ and velocity greater than $10m/s$).  What this in effect achieves, as a result, is a probability measure assignment to a specific target following its description as an algebraic definition of these feature events. In our case, we will study two different datasets. The first one, combines a Radar sensor with an optical sensor. A radar used to explore the velocity of an object among other things, thus defining a sample space and a Sigma-Field with an associated probability measure, will be coupled to a telescopic sensor with also an associated probability space. This product space thus allows us to  define a principled fusion framework where fusing often yields improved and robust performance. Similarly, the second dataset will involve a seismic sensor, coupled with an acoustic sensor.


\section{Background and Related Work} \label{bg}
As noted earlier, sensor fusion has long been of interest, albeit with limited theoretical success particularly when heterogeneous data is present and for which  a unified and systematic approach has remained elusive. An introduction and comprehensive survey to  the area of fusion is provided in \cite{survey1, survey2}. Classical techniques in decision level fusion include Bayesian Inference \cite{dempster2008generalization} and Dempster-Shafer fusion \cite{shafer1976mathematical}. Bayesian Inference has received a lot of attention and has shown success if prior knowledge about sensor reports is available and given.  On the other hand, Dempster-Shafer Fusion was proposed to specifically lift such a restriction on information prior, at a cost of a subtantial  increase in computational complexity. Some applications of Dempster-Shafer Fusion can be found in \cite{rottensteiner2004building}, where, LIDAR data is combined with Multispectral imagery, and in \cite{basir2007engine}, where, multi-sensor information like vibration, sound, pressure, and temperature is fused to detect engine faults. Furthermore, \cite{buede1997target} provides a detailed comparison of Bayesian Inference and Dempster-Shafer Theory. In \cite{huang1993behavior}, a two-stage approach to fusion was proposed, involving knowledge-modeling, which learns from past behavior of the classifiers whose results are to be fused, and operation stage, that combines output of these classifiers based on knowledge learned in the first stage.  More recent work proposes  a sensor model based approach \cite{thesis}, where the sensor network is modeled as either that of  similar or dissimilar sensors. Similar Sensor Fusion or Convex Quadratic Fusion \cite{thesis, li2001convex, florea2007critiques} is used when all the sensors explore the same \topskip 14pt  characteristics/features of the target (for example, a set of 5 radars, looking at the same target), while Dissimilar Sensor Fusion or Analytic Center Fusion \cite{thesis, florea2007critiques} is alternatively used when sensors explore different characteristics/features of the target (for example, a radar and optical sensor looking at the same target). These assumptions turn out to be too restrictive, in that some sensors, albeit dissimilar, may have some common features, while offering additional features to enrich an object/target characterization. Our goal is to explore such a case, and demonstrate that a systematic and principled approach maybe designed, and the overal solution is improved on account of this enhancement. 

\subsection{Mutual Information}
Mutual Information (MI) has been an important and  ubiquitous metric in many applications in data science, with a goal of revealing the synergy/differential between the dependence of two random variables and their lack thereof, by way of their associated probability density functions (PDF).
Specifically, consider two random variables, X and Y, with a joint probability mass $ p(x,y) $ and marginal probability mass functions $ p(x) $ and $ p(y) $. The Mutual Information, $ I(X;Y) $, is the relative entropy between the joint distribution, $ p(x,y) $, and the product distribution, $ p(x)p(y) $ \cite{IT_text} as given by 
\begin{equation} \label{MI}
I(X;Y) = \sum_{x \in X}\sum_{y \in Y} p(x,y)log\frac{p(x,y)}{p(x)p(y)}
\end{equation}
This  relationship may alternatively and perhaps more revealingly,  be expressed in terms of  the variables' joint entropy, their marginal entropies and their  mutual information \cite{IT_text}, 
\begin{equation} \label{MI-JE}
H(X,Y) = H(X) +H(Y) - I(X;Y)
\end{equation}

	 
\section{Problem Formulation} \label{problem formulation}
We assume throughout a set of targets/objects, $ O = \{o_1, o_2, ..., o_I\} $, whose detection and/or classification are of interest. Let the $ k^{th} $ feature observed by the $ l^{th} $ sensor be $ F_k^l $. Then, a set of mutually exclusive events, $ \Omega_k^l = \{a_{kj}^l\}_{j=1, ..., J_{kl}} $, may be defined for the feature $ F_k^l $. Here, $ a_{kj}^l $ is the $ j^{th} $ event for $ F_k^l $ and is described as, $ a_{kj}^l: F_k^l \in [u_j, v_j) $, $ u_j \in  {\rm I\!R^+}$, $ v_j \in {\rm I\!R^+} $, and $ v_j > u_j $. The probability report for the $ k^{th} $ feature from the $ l^{th} $ sensor is then defined as,
\begin{equation}
D_k^l = \{\Omega_k^l, \sigma_B(\Omega_k^l), P_k^l\} 
\end{equation}
Where, $ \sigma_B(\Omega_k^l) $ is the Borel sigma algebra of $ \Omega_k^l $, and can be thought of as the set of all the possible events that can be described over the feature. $ P_k^l $  is the set of probabilities over the events in $ \sigma_B(\Omega_k^l) $.


Since, we define objects as a combination of certain events occurring over different features, we will be working in the product space, 
\begingroup\makeatletter\def\f@size{9.2}\check@mathfonts
\begin{equation}
\Omega = \Omega_1^1 \times \Omega_2^1 \times ... \times {\Omega}_{K_{1}}^1 \times \Omega_1^2 \times \Omega_2^2 \times ... \times {\Omega}_{K_{2}}^2 \times ... \times \Omega_1^L \times \Omega_2^L \times ... \times {\Omega}_{K_{L}}^L
\end{equation}
\endgroup
Where, $ K_l $ is the total number of features observed by the $ l^{th} $ sensor, and $ l = 1, ..., L $. Further, an object will be defined as some combination of events in this product space, $ o_i \in \sigma_B(\Omega)$. Given the object definitions and the probability distributions over various features, we then want to find the fused probability report over the objects, $ D_f = \{O, P_f\} $.

\section{Approach} \label{ap}
The sensor reports form a set  $ \{D_k^l\}_{k=1,\cdots, K_L}^{l=1,\cdots, L} $, 
that come from different sensors, and the target describing events may also be specified over different features. Further, the definitions of specific objects are the result of algebraic operations on the event space $\sigma_B(\Omega)$, a   Sigma-algebra on  the product space, $ \Omega $, with associated probability measures as noted in Section \ref{problem formulation}. Thus, we must determine probability distribution on $\sigma_B(\Omega)$.
\subsection{Determining the joint probability distribution} \label{det joint}

This discussion is in a 2-dimensional setting, in the interest of simplicity, but can be easily extended to an arbitrary number of reports. Let the reports $ D_k^l = \{\Omega_k^l, \sigma_B(\Omega_k^l), P_k^l\}$ and $ D_j^m = \{\Omega_j^m, \sigma_B(\Omega_j^m), P_j^m\} $ be defined over the $ k^{th} $ feature from the $ l^{th} $ sensor, and the $ j^{th} $ feature from the $ m^{th} $ sensor respectively. Then, we have the product space, $ \Omega = \Omega_k^l \times \Omega_j^m $, and we wish to determine a probability distribution  on $\sigma_B(\Omega)$.

When determining the joint probabilities, it is important to  account for the extent of dependence between the features: Completely independent features yield  minimal mutual information, and the joint distribution with minimum mutual information should be selected;  a high correlation between features, on the other hand, yields maximal mutual information, and the joint distribution with maximal mutual information should be selected. These are clearly the extreme cases of dependence, and do not address the partial correlation case. A good approximation to determine the joint probability between partially correlated features  is to consider  a convex combination of the joint probabilities maximizing and minimizing the mutual information. 
The joint probability for the occurrence of events, $\alpha \in \sigma_B(\Omega_k^l)$ and $\beta \in \sigma_B(\Omega_j^m)$, can then be determined as,  
\begin{equation} \label{conv_comb}
P_{\Omega} (\alpha, \beta) =  \rho . {P_{\Omega}}_{\text{MAX MI}} (\alpha, \beta)+ (1 - \rho){P_{\Omega}}_{\text{MIN MI}} (\alpha, \beta),
\end{equation}
where, $ \rho \in [0, 1] $ is a pseudo-measure of extent of correlation between the features. $ \rho \approx 1 $ when the features are highly correlated, and $ \rho = 0 $ when features are independent of each other. $ \rho $ can be determined from the training data, by computing the correlation between the features using a measure like Pearson's correlation or distance correlation \cite{székely2009}.

It can  readily be seen from Equation \ref{MI} that mutual information between two random variables is minimized when the joint probability distribution is selected as the multiplication of the marginal probabilities. Hence, we have,
\begin{equation}
{P_{\Omega}}_{\text{MIN MI}} (\alpha, \beta) = P_k^l(\alpha)*P_j^m(\beta).
\end{equation}
The problem of maximizing mutual information given the marginal probabilities  requires some steps. For some random variables X and Y, conditioning on the marginal probability distributions of X and Y yields constant H(X) and H(Y). As may be seen from Equation \ref{MI-JE}, the maximization of  Mutual Information between two random variables then becomes equivalent to minimizing their Joint Entropy, which is known to be a concave function.
\begin{gather}
\nonumber{P_{\Omega}}_{\text{MAX MI}} = \min_{P_\Omega} \sum_{x \in \Omega_k^l} \sum_{y \in \Omega_j^m} - P_\Omega (x, y) log P_\Omega(x, y)\\
\nonumber\text{subject to: } \sum_{x \in \Omega_k^l} {P_{\Omega}}(x,y) = P_j^m(y),\\	
\sum_{y \in \Omega_j^m}	{P_{\Omega}}(x,y) = P_k^l(x),\text{ } {P_{\Omega}}(x,y) \geq 0. 
\end{gather}
A greedy approach suggested for minimizing joint entropy given the marginal probabilities in \cite{MAXMI} is used in our implementation. The main idea here is to keep large probability masses intact and not break them down into smaller chunks. The contribution of a probability mass toward the joint entropy only increases if it is divided into smaller chunks. That is, for $p = a+b$, $ -p.log(p) \leq -a.log(a) - b.log(b) $, when $ 0<p<1 $ and $ a,b > 0 $. So, keeping the large probability masses from given marginal probabilities intact ensures that their contribution toward the joint entropy is minimized. As empirically demonstrated in \cite{MAXMI}, the minimal joint entropies are obtained to within 1 bit of the optimal values.
\subsection{Determining Object Probabilities} \label{det obj prob}
Once the probability distribution over the product space, $ \Omega $, has been determined, the next step is to determine the object probabilities, $ P_f(o_i) $, in the fusion report, $ D_f = \{ O, P_f \} $. Given the combination of events describing an object, it is pretty straight forward to determine the probability of the presence of an object using the general rules of probability. For an object, o, which is to be described in terms of a combination between $ \alpha \in \Omega_k^l $ and $ \beta \in \Omega_j^m $, we can have either $ o : \{\alpha \wedge \beta\} $ or $o : \{\alpha \vee \beta\} $. Where, $\alpha \wedge \beta $ denotes the combination $ \alpha $ 'and' $ \beta $,  while $ \alpha \vee \beta $ denotes the combination $ \alpha $ 'or' $ \beta $. Then, for the first case we have, $	P_f(o) = P_f(\alpha \wedge \beta) = P_\Omega (\alpha, \beta) $
and for the latter we have, $P_f(o) = P_f(\alpha \vee \beta) = P_k^l(\alpha) + P_j^m(\beta) - P_\Omega (\alpha, \beta) $.
This can be extended to any combination of more than two events.

\section{Experiments and Results} \label{results}

To substantiate the above approach along with the various steps, we select two different datasets. 
\subsection{Dataset 1}
For the first one, we select two sensors, namely a Radar sensor and a telescopic optical sensor, for one (latter) of which we have acquired real data. Due to
 technical difficulty, our Radar measurements were never co-measured with the optical data. Both the sensors are ideally synchronized when observing a given target, which in our case, can be any object in outer space, such as satellite or space debris. For radar simulations we use MATLAB Simulink, and for telescopic image simulation we use some existing telescopic images collected by The Czech Technical University in Prague, and model the object distributions from those images to generate more images with required object characteristics. Each generated radar signal over one second is correlated with two telescopic images. Samples for objects with different velocities, cross-sections, ranges, and aspect-ratios are then generated. The radar signals are used to make decisions over velocity, range, and the cross-section, while the telescopic images are used to make decisions over the aspect-ratio, and displacement over time of an object in view.
\subsubsection{Implementation Details} \label{impl1}
\begin{figure*}[tbp] 
	\centering
	\subfloat[ ]{\includegraphics[width = 0.23\textwidth, height=0.11\textheight]{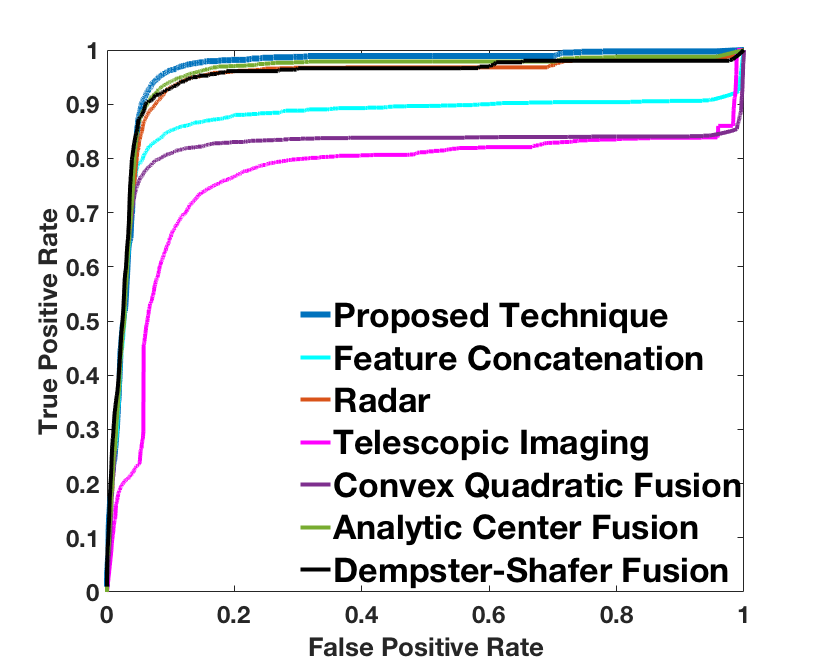}} \hfill 
	\subfloat[ ]{\includegraphics[width = 0.23\textwidth,height=0.11\textheight]{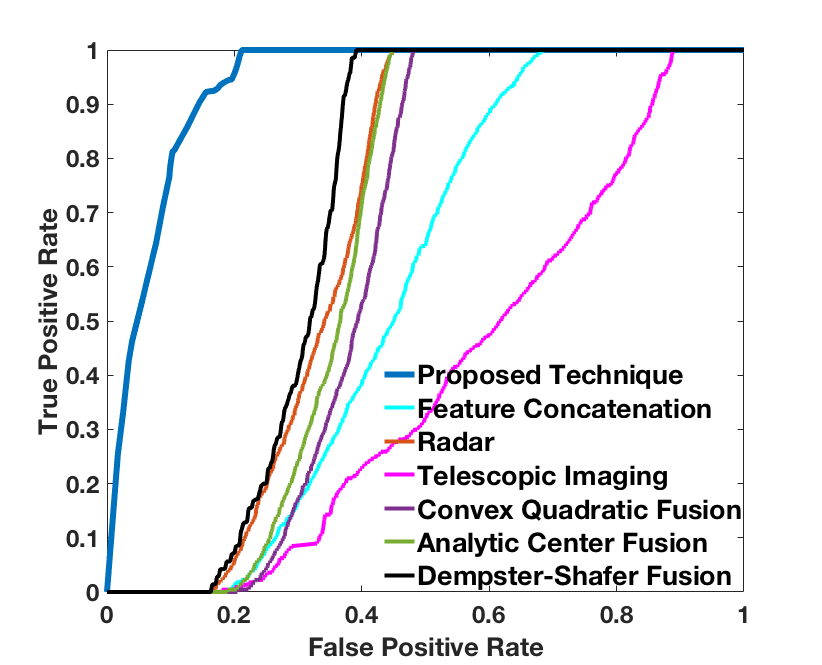}} \hfill
	\subfloat[ ]{\includegraphics[width = 0.23\textwidth,height=0.11\textheight]{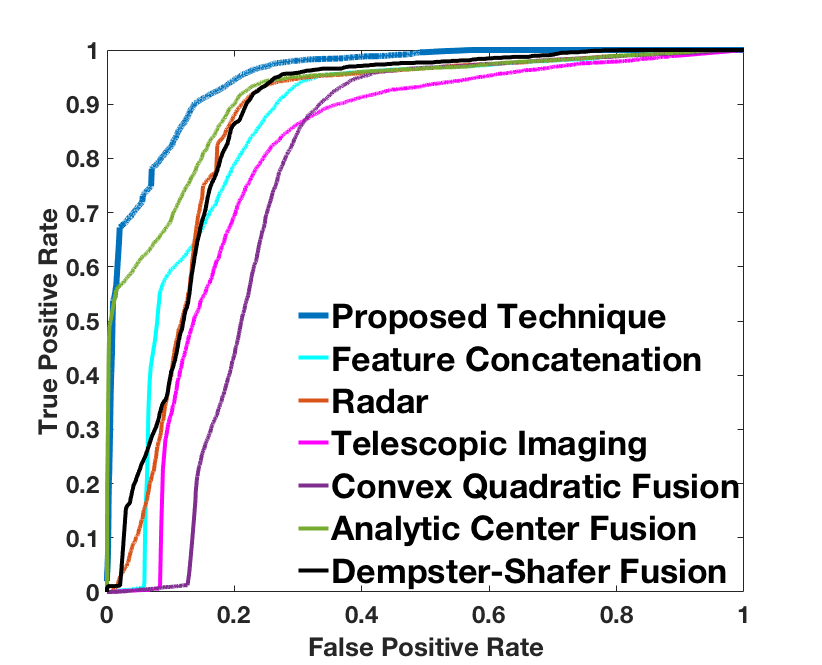}} \hfill
	\caption{\small ROC Curves for detection (Dataset 1) of (a): Class 1:Object 1, (b): Class 2:Object 2, and (c): Class 3:Neither Object 1 nor Object 2}
	\label{roccurves}
\end{figure*}
\begin{figure}[!h]
	\centering
	\includegraphics[width = 0.45\textwidth, height= 0.11\textheight]{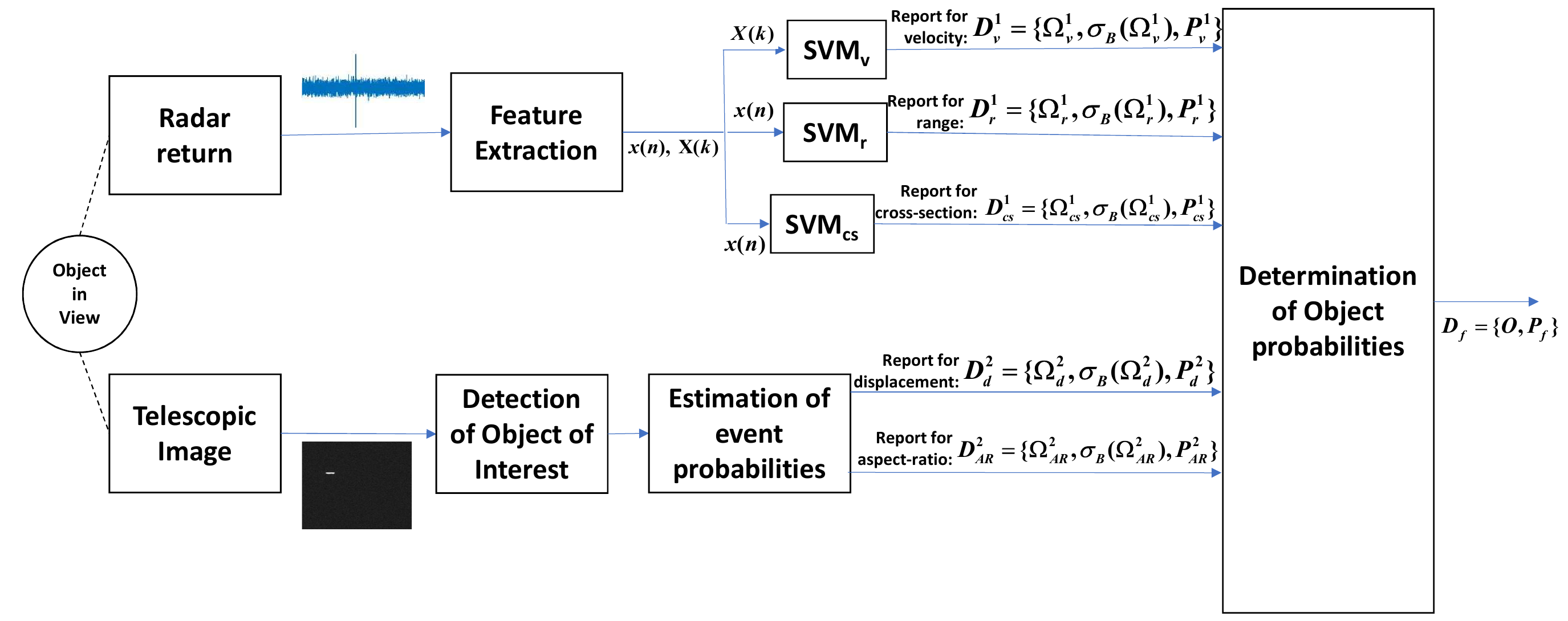}
	\caption{High level block diagram of the implementation, for the first dataset}
	\label{algo}
\end{figure}

To proceed with evaluation, we generate the probability distribution over the features that we will use. Figure \ref{algo} shows a high level block diagram for implementation on the first dataset.

Let the received radar signal be, $ x(n) $, and its corresponding Fourier transform, $ X(k) = \sum_{n=0}^{N-1} x(n)e^{(-i 2 \pi k n/N)} $ for each object, whose labels are distinguished with the velocity, range, and cross-section values of that object, $ [v, r, cs] $. 
Using the training data and labels, SVM classifiers using the Radial Basis Function kernel are trained over the events of interest defined over $[v, r, cs]$, and used to determine a classification score over these events when a test sample is fed in, as shown in Figure \ref{algo}. The classification probabilities are then approximated from the classification score using Platt's technique \cite{Platt}. These probabilities over the events of interest are the, $ P_k^l $, described earlier in Section \ref{problem formulation}.

We have two telescopic images associated with 1-sec of radar return for the same object. Due to space limitations, we forego the details of target detection which is not central to our paper. The object of interest is initially detected by using target detection as discussed in \cite{bian}. Upon its detection, the probability distribution over the object's aspect-ratio, and its displacement in the second image relative to its location in the first image is determined using the image flow technique discussed in \cite{singh1992image}.

\subsubsection{Event and Object Definitions}
For training and testing purposes, we define various events over the feature-sets from both the sensors. For the radar, as noted before, we use $ [v, r, cs] $ and the events are defined as,
\begin{gather}
\nonumber a_1^v: 0 \leq v \leq 10 \text{ }m/s,\text{ } a_2^v: 15\text{ }m/s \leq v \leq 35\text{ }m/s,\\
\nonumber a_1^r: 0 < r \leq 300 \text{ }m,\text{ }a_2^r: 300\text{ }m < r, \\
a_1^{cs}: 0 < cs \leq 20 \text{ }m^2, \text{ }a_2^{cs}: 15\text{ }m^2 \leq cs \leq 50\text{ }m^2.
\end{gather}
From the telescopic imaging sensor, the features, displacement and aspect ratio, $[d, AR]$ define the following events,
\begin{gather}
\nonumber a_1^d: 0 \leq d \leq 60 \text{ }pixels,\text{ }a_2^d: 90\text{ }pixels \leq d \leq 210\text{ }pixels,\\
a_1^{ar}: 0 < AR \leq 1.5, \text{ }a_2^{ar}: 1.5 < AR.
\end{gather}
Furthermore, the objects for classification are defined in terms of these events as,
\begin{gather}
o_1: \{a_1^r \wedge [(a_2^v \wedge a_2^d) \vee (a_2^{cs} \vee a_2^{ar})]\}\\
o_2: \{a_1^v \wedge a_1^d \wedge a_2^r \wedge a_1^{cs} \wedge a_1^{ar}\}
\end{gather}
Given these events and object definitions, we wish to determine the fused report, $ D_f = \{P_f(o_1), P_f(o_2), P_f(\overline {{o_1} \vee {o_2}})\} $. This can be thought of a classification problem with 3 classes, Class 1: Object 1, Class 2: Object 2, and Class 3: Neither Object 1 nor Object 2. 
\subsection{Dataset 2}
The second dataset we used is pre-collected data from a network of seismic sensors, and acoustic sensors deployed in a field, where people/vehicles were walking/driven around in specified patterns. Details about this sensor setup and experiments can be found in \cite{nabritt2015personnel}. This dataset has been previously used for target detection in \cite{lee2017accumulative, sroheda_cmd}, where, the authors focus on detection of humans. Here, we use this dataset to classify between humans, vehicular targets, and no targets. 
\subsubsection{Implementation Details}
\begin{figure}[!h]
	\centering
	\includegraphics[width = 0.45\textwidth, height=0.11\textheight]{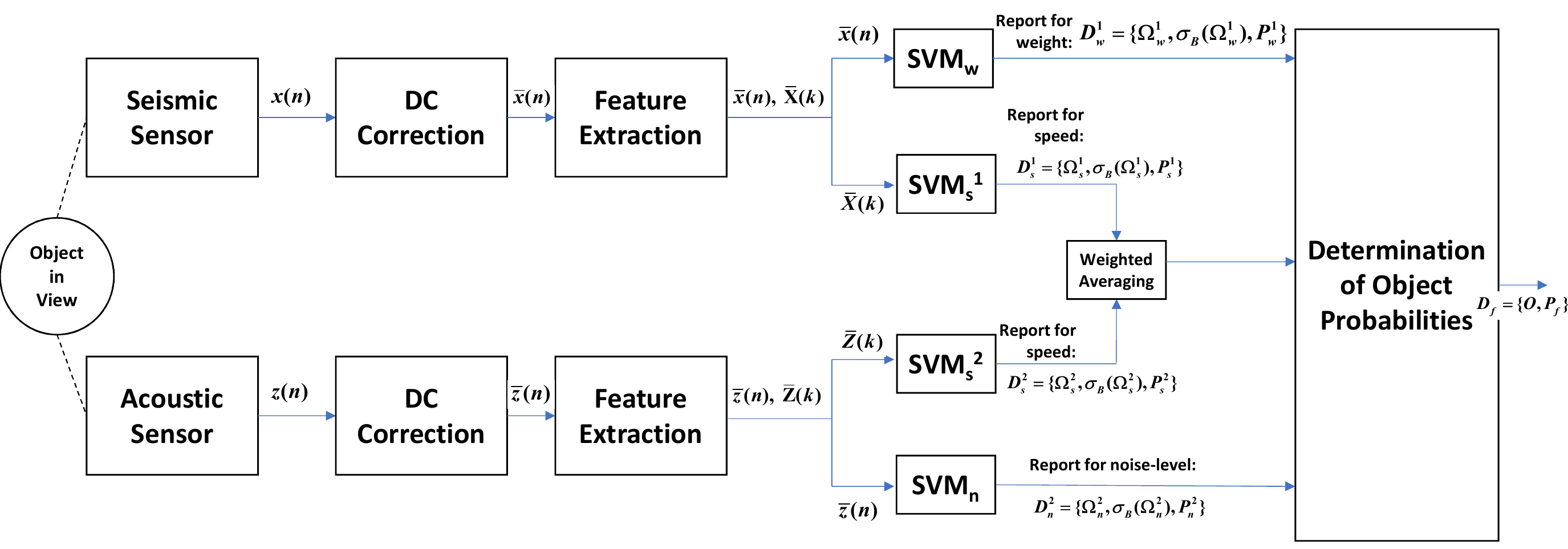}
	\caption{High level block diagram of the implementation, for the second dataset}
	\label{algo2}
\end{figure}

Figure \ref{algo2} shows the high level block diagram for the implementation on the second dataset. 
Using the training data, SVM classifiers are trained over the corresponding events of interest, as discussed before for the first dataset in Section \ref{impl1}. The seismic sensor provides decisions over the features, target weight and target speed, $[w, s]$. True labels for weights of the targets are provided in the dataset, while that for target speed are obtained from the GPS data of the target. Similarly, the acoustic sensor provides decisions on the noise-level of the target, and the target speed, $[n, s]$. The two decisions over the target speed are combined into a single report by performing weighted averaging of the decisions of the two sensors. Here, the weights are selected as per the individual accuracies of the SVMs trained to detect events on target speed.

\subsubsection{Event and Object Definitions}
\begin{figure*}[tbp] 
	\centering 
	\subfloat[ ]{\includegraphics[width = 0.23\textwidth,height=0.12\textheight]{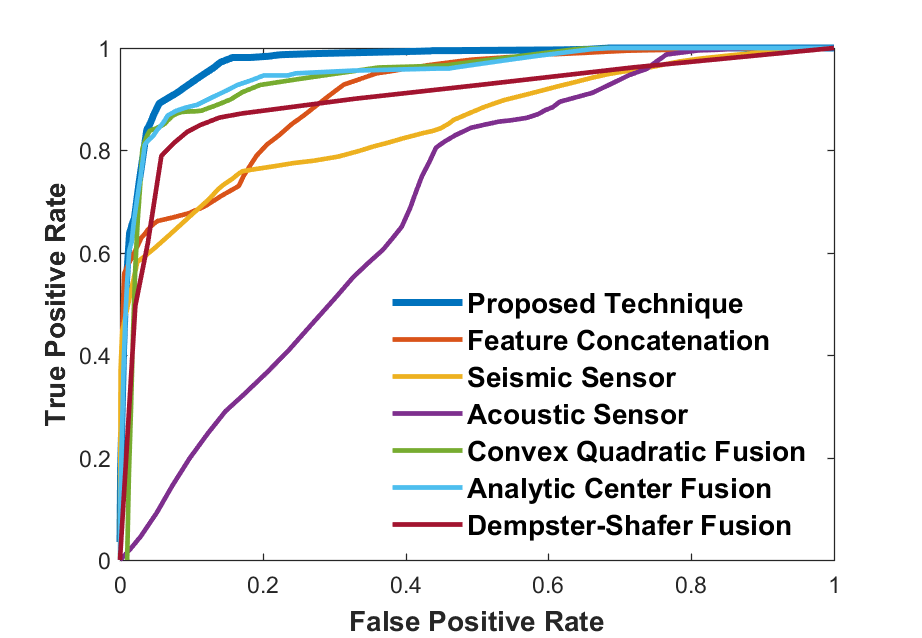}} \hfill
	\subfloat[ ]{\includegraphics[width = 0.23\textwidth,height=0.12\textheight]{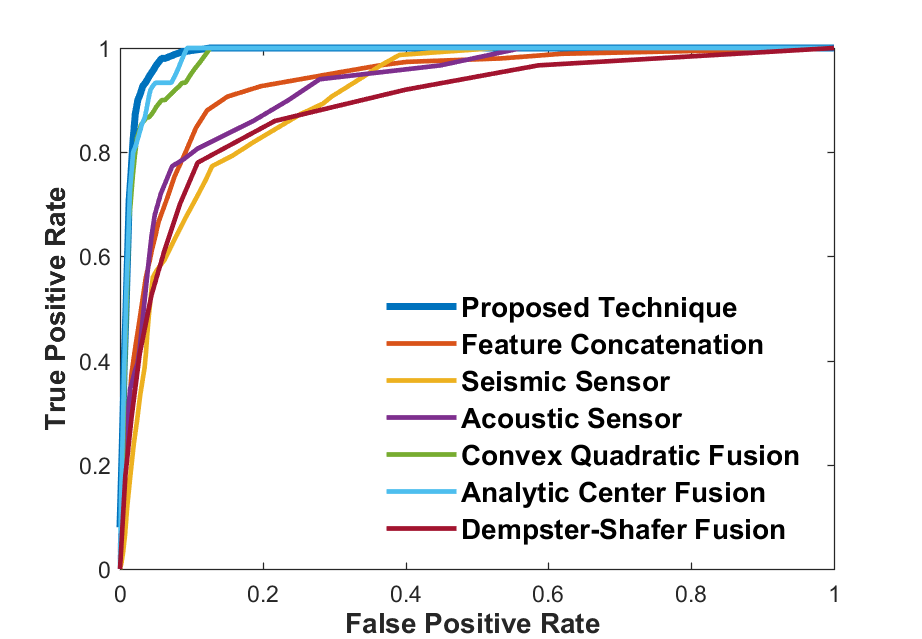}} \hfill
	\subfloat[ ]{\includegraphics[width = 0.23\textwidth, height=0.12\textheight]{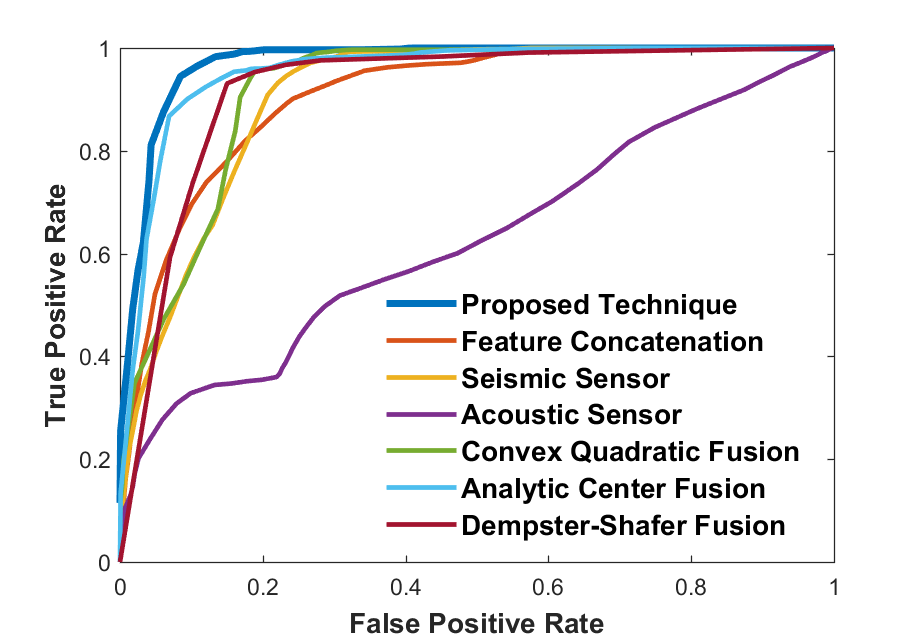}} \hfill
	\caption{\small ROC Curves for detection (second dataset) of (a): Class 1: Human Target, (b): Class 2: Vehicular Target, and (c): Class 3: No Target}
	\label{roccurves2}
\end{figure*}
For training and testing purposes, we define various events over the feature-sets from both the sensors. For the seismic sensor, we use $[w, s]$, while for the acoustic sensor we use $[n, s]$.
\begin{gather}
\nonumber a_1^w: 96.08 \text{ pounds} \leq w \leq 230.61 \text{ pounds},\\
\nonumber a_2^w: 1311.61\text{ pounds} \leq w,\\
\nonumber a_1^s: 0.37 \text{ m/s} < s \leq 2.12 \text{ m/s}, \text{ }a_2^s: 1.7\text{ m/s} \leq s,\\
a_1^n: n \leq -30 \text{ db},\text{ }a_2^n: -10.6658 db \leq n \leq 7.84 db.
\end{gather}
The range of an event can be determined from the training data. The mean of the feature in question over the samples of the same class is computed, and a range of two times the standard deviation is taken on either side of the mean. 
Furthermore, the targets are defined as,
\begin{gather}
	o_1\text{ }(human\text{ }target): \{a_1^s \wedge (a_1^w \vee a_1^n)\}\\
	o_2\text{ }(vehicular\text{ }target): \{a_2^s \wedge a_2^w \wedge a_2^n\}
\end{gather}
Given these events and object definitions, we wish to determine the fused report, $ D_f = \{P_f(o_1), P_f(o_2), P_f(\overline {{o_1} \vee {o_2}})\} $, where, $\{\overline{{o_1} \vee {o_2}}\}$ represents the no target case.
\subsection{Performance Analysis}
Table \ref{acctable}, and \ref{acctable2} show the  classification performance of different techniques (averaged over 10 runs of the technique) when implemented on dataset 1 and 2 respectively.
\begin{table}[htbp]
	\caption{Performance Comparison for the First Dataset}
	\label{acctable}
	\begin{center}
		\begin{tabular}{|c|c|c|c|}
			\hline
			\textbf{Method} & \textbf{Average Accuracy}\\
			\hline
			Radar & 86.47\%\\
			\hline
			Telescopic Imaging & 81.31\%\\
			\hline
			Feature Concatenation & 85.93\%\\
			\hline
			Convex Quadratic Fusion (Similar Sensor Fusion)  & 86.07\%\\
			\hline
			Analytic Center Fusion (Dissimilar Sensor Fusion)  & 88.61\%\\
			\hline
			Dempster-Shafer Fusion & 87.18\%\\
			\hline
			\textbf{Proposed Technique} & \textbf{90.36\%}\\
			\hline
		\end{tabular}
	\end{center}
\end{table}

\begin{table}[htbp]
	\caption{Performance Comparison for the Second Dataset}
	\begin{center}
		\begin{tabular}{|c|c|c|c|}
			\hline
			\textbf{Method} & \textbf{Average Accuracy}\\
			\hline
			Seismic Sensor & 85.41\%\\
			\hline
			Acoustic Sensor & 67.62\%\\  
			\hline
			Feature Concatenation & 81.63\%\\
			\hline
			Convex Quadratic Fusion (Similar Sensor Fusion)  & 86.69\%\\
			\hline
			Analytic Center Fusion (Dissimilar Sensor Fusion) & 89.96\%\\
			\hline
			Dempster-Shafer Fusion & 87.93\%\\
			\hline
			\textbf{Proposed Technique} & \textbf{92.04\%}\\
			\hline
		\end{tabular}
		\label{acctable2}
	\end{center}
\end{table}
Classification accuracy is often not the best measure to quantify performance, particularly in cases where different classes have different numbers of samples, which is the case here. A better way to compare performance is to look at the Receiver Operating Characteristic (ROC) curves. Fig. \ref{roccurves} and \ref{roccurves2} show the ROC curves for classification in case of datasets 1 and 2 respectively. It can be seen from the ROC curves (for dataset 1) in Fig. \ref{roccurves} that other techniques show limited performance in correct classification of objects from class 2 due to the low number of samples for class 2 in comparison to those in class 1 and class 3. This causes the classifier to bias toward selecting class 1 or class 3, even when the sample is from class 2 in order to achieve high classification accuracy. But, our technique trains over occurrence of events rather than the object itself, hence does not face this issue. 
Improvement in performance is also seen for Dataset 2 (Fig. \ref{roccurves2}). In particular, detection of human targets is significantly improved, by taking 'or' between noise level event and weight event, which reduces misclassification due to noise of winds. 

\section{Conclusion}
We propose a novel sensor fusion technique at the decision
level that looks at targets as combinations of events over the
features that describe it, while also considering the extent
of correlation between different features. Experiments on various datasets showed that the proposed technique can outperform existing fusion techniques on the decision
level. 

\end{document}